\documentclass[10pt,twocolumn,twoside]{IEEEtran}
\usepackage[pdftex]{graphicx}
\newsavebox\mybox
\usepackage{color}
\usepackage{cite}
\usepackage{amsmath,amssymb}
\usepackage{siunitx}
\usepackage{multirow}
\usepackage{dblfloatfix}
\usepackage{xcolor}
\usepackage{subcaption}

\usepackage{amsthm}
\newtheorem{theorem}{Theorem}

\newtheorem{remark}{Remark}
\newtheorem{lemma}{Lemma}
\newtheorem{definition}{Definition}

\begin{document}
\renewcommand{\arraystretch}{1.15}

\title{Stability of DC Networks with Generic Load Models}

\author{Kathleen Cavanagh, Petr Vorobev, Konstantin Turitsyn

\thanks{Authors are with MIT Department of Mechanical Engineering,  Cambridge, MA, 02139}}

\maketitle

\begin{abstract}
DC grids are prone to small-signal instabilities due to the presence of tightly controlled loads trying to keep the power consumption constant over range of terminal voltage variations. Th, so-called, constant power load (CPL) represents a classical example of this destabilizing behavior acting as an incremental negative resistance. Real-life DC loads represented by controlled power converters exhibit the CPL behavior over a finite frequency range. There exist a number of methods for stability certification of DC grids which are primarily concerned with the source-load interaction and do not explicitly account for the influence of network. In the present manuscript, we develop a method for stability assessment of arbitrary DC grids by introducing the \emph{Augmented Power Dissipation} and showing that it's positive definiteness is a sufficient condition for stability. We present an explicit expression for this quantity through load and network impedances and show how it could be directly used for stability certification of networks with arbitrary configuration. 
\end{abstract}


\IEEEpeerreviewmaketitle


\section{Introduction}\label{sec:introduction}

Recent advances in power electronics technologies and general trend towards renewable energy sources have lead to increased interest in DC grids \cite{planas2015ac,justo2013ac}. Small-scale DC microgirds have been in use for several decades already, mainly as autonomous electric systems on board of vehicles \cite{emadi2003vehicular}. As such, the configuration of these microgrids was fixed and well planned for the exact operating conditions which were known in advance. On the other hand, DC microgrids with "open" structure, capable of being expanded and reconfigured, as well as suitable for stable operation in a broad range of loading conditions, have mostly been out of the scope of academic research. In part, this is justified by the fact that, currently, the majority of microgrids are using AC interface for power distribution, even if all the sources and loads are naturally DC. However, fully DC microgrids can become an economically feasible solution for supplying power in remote communities at the minimal cost.

Unlike in AC grids where substantial part of the load is of electro-mechanical type, loads in DC grids are mostly represented by power electronics converters with tight controls to achieve flat voltage at their outputs \cite{erickson2007fundamentals}. This leads to a constant power load (CPL) behavior on the input within the control loop bandwidth, which is regarded to be one of the main sources of instabilities in DC grids. The origin of the instability is often regarded to the, so-called, negative incremental resistance introduced by CPL and a number of methods for stability assessment are based on such representation. Recent reviews \cite{singh2017constant,riccobono2012comprehensive}  present a comprehensive classification of the existing stability criteria and stabilization methods for DC grids. 

Apart from certain special configurations \cite{cupelli2015ideal} CPL provides the worst case representation for real-life converters, formally corresponding to constant negative incremental resistance over all the frequencies. Recently the problem of stability of CPL based microgrids has attracted substantial attention from the controls community, see e.g. \cite{Sanchez:2013gl, SimpsonPorco:2015hp, Zhao:2015eu, Barabanov:2016ki, de2018power}.
In our previous works on the subject \cite{belk2016stability, inam2016stability, Cavanagh:2017uo} we have demonstrated that the problem of linear and transient stability of DC microgrids with CPL can be addressed using Brayton-Moser mixed potential approach \cite{Brayton:1964gr, Jeltsema:2009jd, SchaftVanDer:2014wo}. However stability conditions derived under the CPL modeling assumptions may potentially be conservative and lead to excessive constraints on network configuration and/or installed equipment. In reality, regulated power converters act similar to CPL - exhibiting incremental negative resistance properties - only within their control loop bandwidth, at frequencies lower than the cross-over frequency \cite{rahimi2009analytical}. Above this frequency the effective incremental resistance is positive and no special stabilization is required so that less conservative stability criteria can be formulated.    

Traditionally, power electronics community has relied on a number of different impedance based stability conditions. Most of them consider the minor loop gain which is the ratio of the source output impedance to the load input impedance. The celebrated Middlebrook criterion \cite{MB1976}, originally proposed for input filter design, is based on a small-gain condition for minor loop gain demanding its absolute value to be less than unit (or even less if a gain margin is imposed) for all frequencies. It is a rather conservative method, however it requires only the knowledge of the absolute values of impedances over the whole frequency range. A less conservative Gain Margin Phase Margin criterion \cite{wildrick1995method} allows the Nyquist plot of the minor loop gain to leave the unit circle provided there is a sufficient phase margin. Another method - the opposing argument criterion - is based on conditions imposed on the real part of the minor loop gain \cite{feng2002impedance}. The main advantage of this method is that it can be applied to multiple load systems, since the contributions from each individual load can now be simply added together. Finally, the least conservative methods are the, so-called, energy source consortium analysis (ESAC) \cite{sudhoff2000admittance} and a similar one root exponential stability criterion \cite{sudhoff2011advancements}. Both offer the smallest forbidden region for the minor loop gain of all the existing methods. 

Complementary to stability assessment methods there exist a number of stability enhancement techniques aimed at increasing of the effective system damping. This can be either by adding/changing the circuit elements - passive damping,, or by changing converters control loops - active damping \cite{singh2017constant}. In the present manuscript we are interested in the former considering both source and load control settings to be fixed. Passive damping is realized by adding circuit elements to load input (or modifying the input filter) in order to reshape the load input impedance. The choice of different damping circuits for stabilization of a CPL is analyzed in \cite{cespedes2011constant} while an optimal choice of damping resistance is given in \cite{xing2011optimal}. 

The stability assessment and enhancement methods reviewed above a mostly concerned with single-source single-load systems and a possible destabilizing interaction between the impedances of  corresponding components. Under certain conditions it could be possible to generalize some of these methods by attributing the network parts to either loads or sources effectively changing corresponding impedances. However, there is a need for new methods for stability assessment which can be routinely applied to networked DC microgrids where the effect of line impedances on system stability is substantial. 

The key contribution of this manuscript is a novel stability certification method based on a specially formulated passivity condition for the whole system. We quantify the destabilizing effect of the loads by introducing a certain function - \emph{Augmented Power Dissipation} - and show that the positive definiteness of this function represents a sufficient stability condition. We describe a procedure for calculating the contribution to the Augmented Power Dissipation from loads and network and present a method for splitting the network contribution into parts each responsible for stabilizing of a single load.


    
    
    


\section{Problem Formulation}\label{sec:formulation}

DC-DC power electronic converters are typically designed to efficiently provide a fairly constant output voltage with control system properly mitigating the input voltage variations. As seen from the network these converters represent a rather complex dynamic behavior, so a dynamic model of the converter needs to be considered for studying the stability of it's operation with the network. A common way to represent a lossless converter is using a single-pole, double-throw (SPDT) switch with a duty cycle $D$ defining the fraction of the cycle spent in position 1. By using lossless elements, the modification of the input voltage is achieved in a manner that can apply to all power throughputs. In steady state, converters can be described by their DC conversion ratio $M$, the ratio of the output voltage $V$ to the input voltage $V_g$. This conversion ratio is a function of the duty cycle control input.

To study stability, we utilize a small signal model of the system. Due to the switch, which operates with a frequency on the order of tens of kilohertz, the inductor current and capacitor voltage vary throughout a given cycle with small amplitude around their average values even in steady state. Therefore, to properly account for the low frequency variations, the so-called small-ripple approximation is used where the system is averaged over a period so the switching variations within a given period are neglected. The averaged system can then be linearized about a given operating point to give a small signal model. 

\begin{figure}
  \centering
  \begin{subfigure}[t]{\linewidth}
    \centering
    \includegraphics[height= 0.22\linewidth]{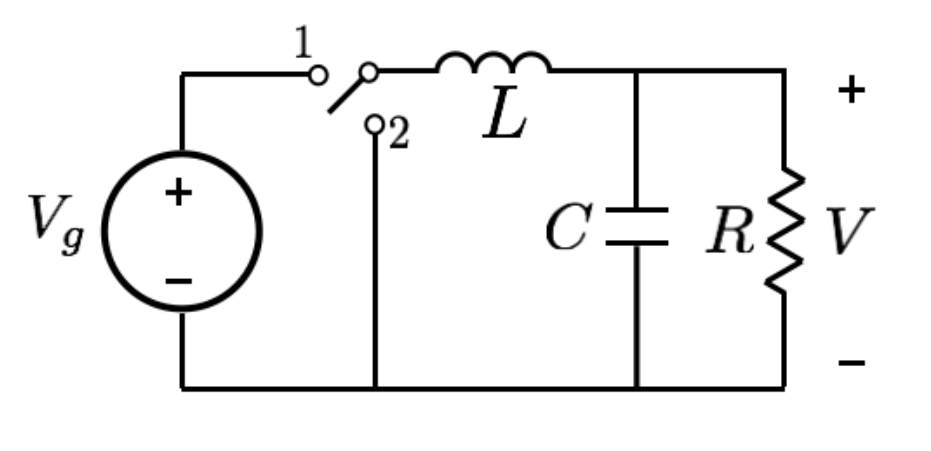}
    \caption{Switching Model}
    \label{fig:buck}
  \end{subfigure}\\
  \centering
    \begin{subfigure}[t]{\linewidth}
    \centering
    \includegraphics[height =0.22 \linewidth]{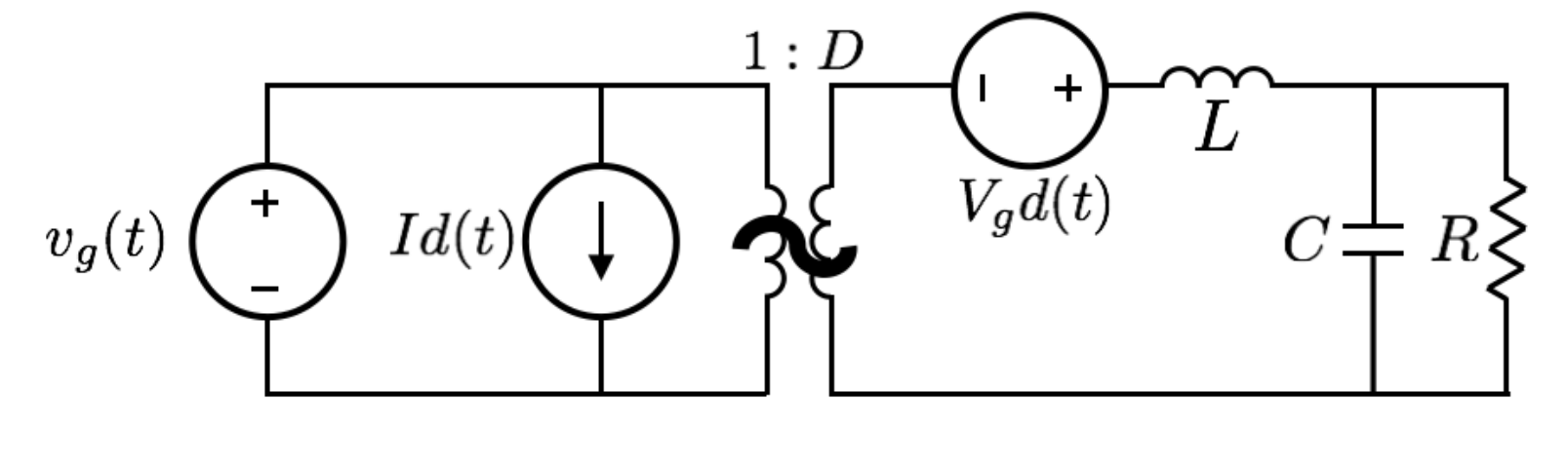}
    \caption{Small signal model}
    \label{fig:buck_SS}
  \end{subfigure}
  \caption{Steady state and small signal models of a buck converter}
\end{figure}

While there are several common types of converters, we will focus on the buck converter,
given in Figure \ref{fig:buck}, due to its prevalence. The buck converter, consisting of an inductance $L$, capacitance $C$ and constant resistance load $R$, has a conversion ratio $M(D) = D$, such that the converter steps-down the input voltage in proportion to the duty cycle. The small signal model of a buck converter is shown in Figure \ref{fig:buck_SS} where all uppercase variables refer to the operating point and all lowercase variables refer to the time varying deviations. Since averaging is done over many switching periods, an ideal DC transformer, rather than a switch is used in the small signal model to properly describe dynamics. A formal derivation of this model and additional background on buck converters are given in \cite{erickson2007fundamentals}. 

\begin{figure}
    \centering
    \includegraphics[ width=0.9\linewidth]{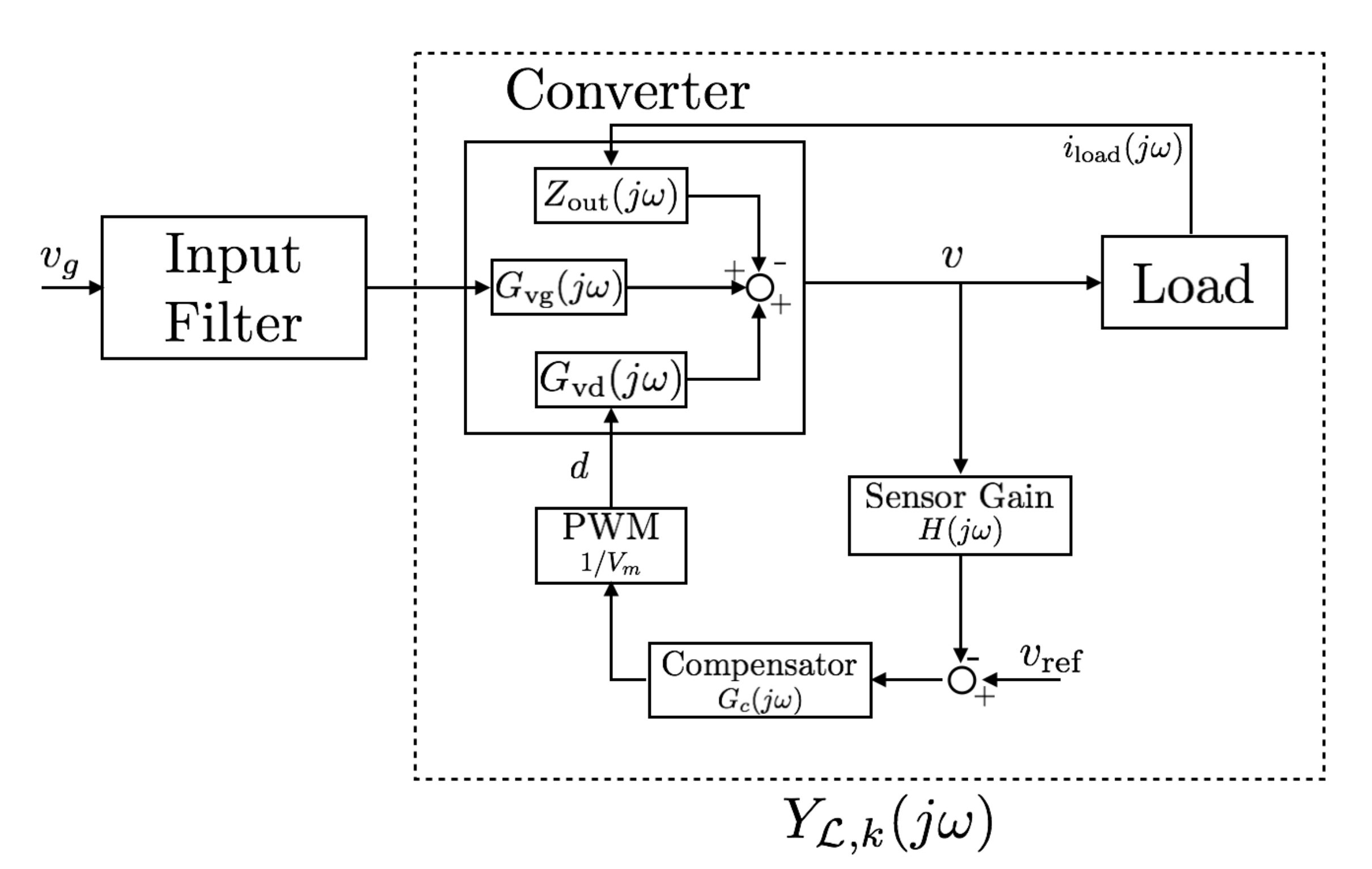}
    \caption{Definition of Small Signal Load Model}
    \label{fig:Load_Model}
\end{figure}

To maintain constant voltage at its output, a converter relies on feedback to modify the duty cycle following any variations in the input voltage. In this work, we consider a buck converter interfaced with a constant resistance on it's output and the control loop to stabilize the output voltage, so that it is seen from the network as a constant power load in steady state. This model is defined in Figure \ref{fig:Load_Model}. To characterize the small signal behavior of the buck converter, we first examine the transfer function from the control input (duty cycle variation) $d(s)$ to the output voltage $v(s)$ which has the following form (detail can be found in \cite{erickson2007fundamentals}) 

\begin{equation}
    G_{vd}(s) = \frac{v(s)}{d(s)} = G_0\frac{\omega_0^2}{s^2+2\zeta\omega_0s+\omega_0^2}
\end{equation}
with $G_0 = V/D$, $\omega_0 = 1/\sqrt{LC}$ and $\zeta = 1/(2R)\sqrt{L/C}$. In the presence of voltage control, the loop gain of the system is defined as $T = G_{vd}(s)H(s)G_c(s)/V_m$ where $H$ is the sensor gain, $G_c(s)$ is the compensator transfer function and $V_m$ is the voltage of the pulse width modulation. The input admittance transfer function $Y_{\mathcal{L}, k}(s)$ can then be defined as 

\begin{equation}
    Y_{\mathcal{L}, k}(s) =\frac{i(s)}{v_g(s)} = \frac{1}{Z_N(s)}\frac{T(s)}{1+T(s)}+\frac{1}{Z_D(s)}\frac{1}{1+T(s)}
\end{equation}
where $Z_N$ is the converter impedance when the output $v(s)$ is nulled and $Z_D$ is the converter impedance in the absence of control input $d(s)$. For a buck converter, these have the following expressions: $Z_N=\frac{-R}{D^2}$ and $Z_D=\frac{R}{D^2}\frac{1+sL/R+s^2LC}{1+sRC}$. 

\begin{figure}
    \centering
    \includegraphics[width=.6\linewidth]{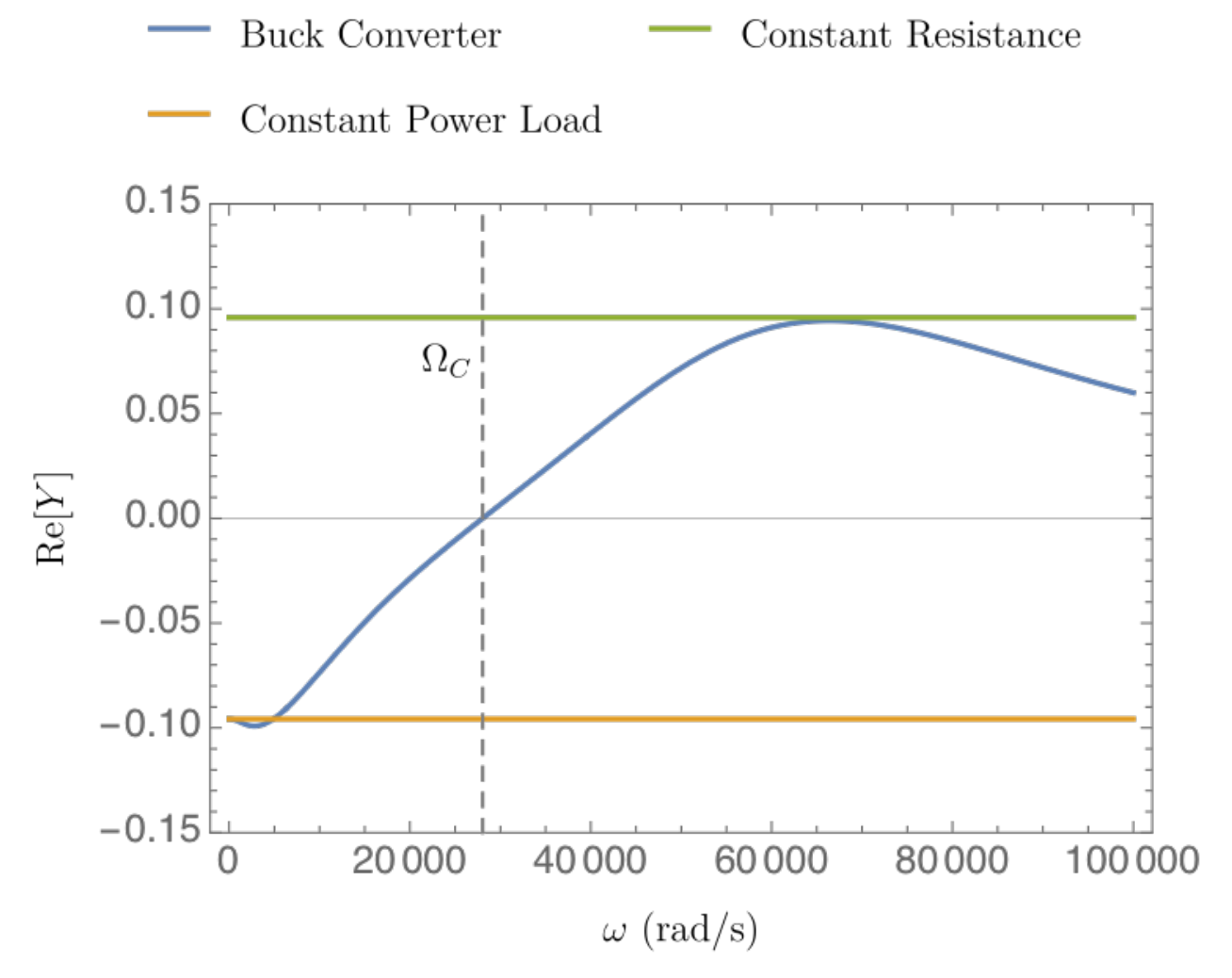}
    \caption{Real component of the input admittance transfer functions for a buck converter, constant resistance load and constant power load.}
    \label{fig:BuckTF}
\end{figure}

 As will be shown in the following sections, much of the analysis relies on the real component of the input admittance. For a buck converter, this quantity is given in Figure \ref{fig:BuckTF}, the corresponding quantities for constant resistance load and a constant power load are given for comparison. A tightly controlled buck converter can be thought of as a constant power load at low frequencies. The buck converter becomes passive at a crossover frequency $\Omega_C$, defined as the frequency above which $\text{Re}[Y_\mathcal{L}(\omega)]>0$. In contrast the constant power load is frequency independent and is never passive. This highlights the need for a more detailed stability analysis than a constant power representation can provide for systems with power converters.  

In practice, converters are also equipped with an input filter to reject the high-frequency current components from converter switching actions. In this work, the input filter is considered to be part of the network. It is known that improper input filter design can bring instability to an otherwise stable converter, therefore, a careful selection of filter configuration is required. Typically, the whole system is designed to be stable when directly connected to a fixed DC voltage source with the effect of interconnecting line neglected. If the connecting line has substantial impedance, it can interfere with the converter filter and finally lead to instability.


\section{Effective Load Admittance and Stability Conditions}\label{sec:genap}

\subsection{Effective Admittance Representation}

To study small-signal stability of our network we introduce the following vector of node voltage and current perturbations  $x_e=[i_e^T. v_e^T]^T$ where both the node current $i \in \mathbb{C}^{|\mathcal{V}|}$  and node voltage $v \in \mathbb{C}^{|\mathcal{V}|}$ being deviations from the
equilibrium point, eg. $i_e(t)=I(t)-I_0$.

Next, we define the frequency dependant nodal admittance matrix according to a general rule:

\begin{equation}
    i(\omega)=Y(\omega) v(\omega)
\end{equation}

\noindent where the full matrix Y includes contribution from network   $Y_\mathcal{N}$  and loads $Y_\mathcal{L}$ with the latter matrix being diagonal. The effective load admittance matrix can be explicitly expressed using converter small-signal model or directly measured.

Our general strategy for proving the stability of the interconnection is based on the well-known ``zero exclusion principle'' and is closely related to the $\mu$-analysis.
\begin{theorem}
 The interconnection is stable for any admissible load transfer functions if there exists a quadratic form
\begin{equation} \label{eq:GenericDissipation}
 W = \mathrm{Re}[v^\dag D (Y_\mathcal{N}(\omega) + Y_\mathcal{L}(\omega)) v]
\end{equation}
that is strictly positive for any harmonic signal $v(\omega) \neq 0$.
\begin{proof}
 This is a well-known principle in the robust control literature. The simple interpretation of the standard proof is the following. Suppose that for some admissible loads there exists a pole in the right hand plane. Consider now a homotopy between the unstable an completely unloaded system (all load admittance are zero $Y_\mathcal{L} = 0$). As the unloaded system is stable, there would be a point on a homotopy where the originally unstable pole lies on the imaginary axis at $\omega = \omega_c$. The voltage profile $v$ corresponding to this mode, would then satisfy $(Y_\mathcal{N}(\omega_c) + Y_\mathcal{L}(\omega_c)) v = 0$ and contradict the assumptions of the theorem.
\end{proof}
\end{theorem}

In the following we consider a specific quadratic functional we refer to as \emph{Augmented Power Dissipation}, that is characterized by $D = \mathrm{e}^{j\phi(\omega)}$. The important property that follows directly from \eqref{eq:GenericDissipation} is that the total dissipation of the system can be decomposed into the contributions coming from the network $W_\mathcal{N} = \mathrm{Re}[v^\dag D Y_\mathcal{N}(\omega) v] $ and loads $W_\mathcal{L} = \mathrm{Re}[v^\dag D Y_\mathcal{L}(\omega) v] $ and one can establish the bounds on these components separately. After proving some intermediate results, we derive the sufficient conditions on the network structure that guarantee overall system stability.
\begin{definition}
The Augmented Conductance of a two-terminal element with admittance $y_k$ is defined as
\begin{equation}
 G_k(\omega, \phi(\omega)) = \mathrm{Re}[\mathrm{e}^{j\phi(\omega)} y_k(\omega)]
\end{equation}
\end{definition}
\begin{definition}
 For a two terminal element characterized by the admittance $y_k(\omega)$ and voltage drop $\nabla_k v(\omega)$, the Augmented Power Dissipation is defined as
\begin{equation}
 W_k = \mathrm{Re}\left[\overline{\nabla_k v}\mathrm{e}^{j\phi} y_k\nabla_k v\right] = G_k(\omega)|\nabla_k v|^2
\end{equation}
\end{definition}
The Augmented Power Dissipation coincides with real power dissipation for $\phi = 0$ and with reactive power dissipation for $\phi = \pi/2$. For general $\phi(\omega)$ the corresponding time-domain storage function does not have any well-defined physical meaning. The following results will be  useful for constructing the dissipative functionals:

\begin{remark}\label{remark:Inductive}
The Augmented Conductance of resistive-inductive elements with admittance $y = 1/(r + j\omega L)$ is given by
\begin{equation}\label{eq:InductiveG}
 G = \frac{r \cos\phi(\omega) +\omega L \sin\phi(\omega)}{r^2 + \omega^2 L^2}
\end{equation}
The Augmented Conductance is positive whenever $r\cos(\phi(\omega))+\omega L \sin(\phi(\omega))>0$. The biggest value of Augmented Conductance, and thus the highest rate of Augmented Dissipation is achieved for $\tan\phi(\omega ) = {\omega L/ R}$  for which $G =1/ \sqrt{r^2 + \omega^2 L^2}$
\end{remark}

\begin{remark}\label{remark:Capacitive}
For resistive-capacitive elements with $y = 1/(r-j/(\omega C))$ the Augmented Conductance takes the form
\begin{equation}\label{eq:CapacitiveG}
 G = \omega C\frac{ \omega r C \cos\phi(\omega) - \sin\phi(\omega)}{1+ (\omega r C)^2}
\end{equation}
This element remains augmented dissipative whenever $r\omega C \cos(\phi(\omega)-\sin(\phi(\omega))>0$
\end{remark}

\begin{remark}\label{remark:CapacitiveShunt}
In the presence of shunt capacitive element with inductive and resistive line with $y = 1/(r+j\omega L)+j\omega C$ the augmented conductance takes the form
\begin{equation}\label{eq:CapacitiveShunt}
 G =   \frac{r \cos\phi(\omega) +\omega L \sin\phi(\omega)}{r^2 + \omega^2 L^2} - \omega C \sin \phi(\omega) 
\end{equation}
A choice of negative $\phi$ may be required for this element to remain augmented dissipative.  
\end{remark}

Whenever the overall element admittance magnitude is bounded, the following bound can be established on the absolute value of Augmented Power Dissipation
\begin{lemma} \label{lemma:Ybound}
For any two-terminal element $k$, with bounded magnitude of admittance $y_k$ satisfying $|y_k| \leq Y^{\max}_k$, the absolute value of Augmented Power Dissipation is bounded from above:
\begin{equation}
    \left|\mathrm{Re}\left[\overline{\nabla_k v}\mathrm{e}^{j\phi} y_k \nabla_k v\right]\right| \leq Y^{\max}_k |\nabla_k v|^2
\end{equation}
\begin{proof}
Follows directly from $|\mathrm{Re}[\mathrm{e}^{j\phi} y_k]| \leq |\mathrm{e}^{j\phi} y_k| = |y_k|$
\end{proof}
\end{lemma}
This observation implies that the constant power load with incrementally negative resistance can inject only some bounded amounts of Augmented Power into the system. Our stability certificates will be based upon matching the Augmented Power injected by the loads with it being dissipated by the network elements.

\subsection{Network Decomposition and Stability Certificates}

In order to derive the stability certificate, we first introduce a path based characterization of the interconnection network.
\begin{definition}\label{def:load_path}
A load path $\Pi_{k}(\omega) = \{e_1,\dots e_{n_k}\}$ is an ordered set of directed circuit network edges satisfying following properties: i) the end of edge $e_m$ coincides with the beginning of edge $e_{m+1}$  ii) the beginning of the edge $e_1$ coincides with the load $k \in\mathcal{L}$, iii) no edge appears twice in the path, iv) the end of the last edge $e_{n_k}$ does not coincide with the origin and is denoted as $\mathrm{end}(\Pi_k)$, and v) the augmented conductance $G_e$ of every element $e\in\Pi_k$ along the path is positive: $G_e >0$.
\end{definition}
Note, that explicit dependence of the path $\Pi_{k}(\omega)$ on the frequency empasizes that different paths can be used for different frequencies. For every path we define the aggregate Augmented Conductance:
\begin{definition}
The aggregate Augmented Conductance $G_{\Pi_k}$ of the path $\Pi_k$ is related to the conductance of individual elements via the following series interconnection relation:
\begin{equation} \label{eq:PathConductance}
 \frac{1}{G_{\Pi_k}} = \sum_{e\in \Pi_k} \frac{1}{G_e}
\end{equation}
\end{definition}
This definition appears naturally in the following bound on the Augmented Power Dissipation:
\begin{lemma}\label{lemma:PathPower}
 If the network is composed of elements with positive Augmented Conductances $G_k > 0$, the aggregate Augmented Power Dissipation $W_{\Pi_k} = \sum_{e\in\Pi_k} W_e$ of all elements $e \in \Pi_k$ along the path is bounded from below by
\begin{equation}
 W_{\Pi_k}  \geq G_{\Pi_k}\left|v_k - v_{\mathrm{end}(\Pi_k)}\right|^2
\end{equation}
\begin{proof}
 Define $\pi_e = G_{\Pi_k}/G_e > 0$, so that $\sum_{e\in \Pi_k} \pi_e  = 1$ in accordance to \eqref{eq:PathConductance}. In this case one has
\begin{align}
 W_{\Pi_k}  & = \frac{1}{G_{\Pi_k}}\sum_{e\in\Pi_k}\pi_e \left|G_e\nabla_e v\right|^2 \nonumber\\
 &\geq  \frac{1}{G_{\Pi_k}}\left|\sum_{e\in\Pi_k}\pi_e G_e\nabla_e v\right|^2 \nonumber\\
 &= G_{\Pi_k}\left|v_k - v_{\mathrm{end}(\Pi_k)}\right|^2
\end{align}
Here we have used the Jensen's inequality and the definition of the path.
\end{proof}
\end{lemma}
\begin{definition}
A load path decomposition of the network is a set of load paths $\Pi_k$ assigned for every load $k \in\mathcal{L}$, such that there is only one path passing through every two-terminal element $k\in\mathcal{E}$.
\end{definition}

\savebox{\mybox}{\includegraphics[width=0.25\linewidth]{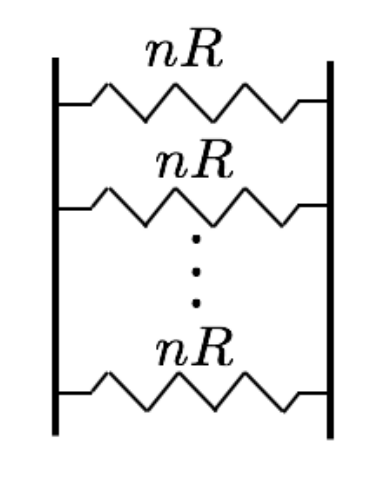}}
\begin{figure}
    \centering

    \begin{minipage}{0.25\linewidth}
        \centering
        \subcaptionbox{}{\vbox to \ht\mybox{%
            \vfill
            \includegraphics[width=\linewidth]{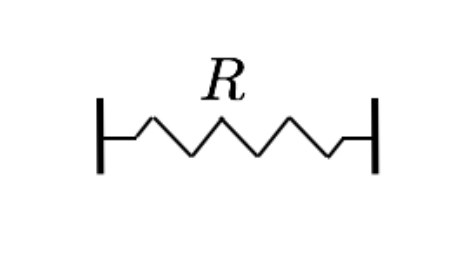}
            \vfill
        }}
    \end{minipage}
    \begin{minipage}{0.25\linewidth}
        \centering
        \subcaptionbox{}{\usebox{\mybox}}
    \end{minipage}

    \caption{The example decomposition of a two port resistive element to allow for $n$ paths.}
    \label{fig:path_decomp}
\end{figure}

 Note, that the load path decomposition may not exist for original circuit, for example there is only a single source  and multiple loads connected to it through a single line. However, 
 the path decomposition can be constructed for an equivalent electric circuit where some individual network elements are represented as a parallel interconnection of multiple ones, like illustrate on Figure \ref{fig:path_decomp} . In the following, we assume that the network circuit has been transformed to allow for load path decomposition. Then, the following lemma establishes the ground for the central result of the paper:
\begin{lemma}\label{lemma:NetworkPower}
The total Augmented Power $W_{\mathcal{N}}$ dissipated in the network characterized by the path decomposition $\Pi_{k}$ with $k \in \mathcal{L}$ is bounded from below by the following expression:
\begin{equation} \label{eq:NetworkPowerBound}
 W_{\mathcal{N}} \geq \sum_{k}G_{\Pi_k}\left|v_k - v_{\mathrm{end}(\Pi_k)}\right|^2
\end{equation}
\begin{proof}
Given that there is at most one path passing through every element, one has $W_{\mathcal{N}} \geq \sum_{k\in\mathcal{L}} W_{\Pi_k}$. Then the relation \eqref{eq:NetworkPowerBound} follows immediately from Lemma \ref{lemma:PathPower}.
\end{proof}
\end{lemma}
This allows us to finally formulate the following central result of the work:
\begin{theorem} \label{thm:final_result}
Consider a network composed of elements with positive Augmented Conductances and constant voltage sources with $v_k = 0$. Assume that for every $\omega$, there exists a path decomposition $\Pi_k(\omega)$ such that every path ends either at the ground or source nodes, i.e. $v_{\mathrm{end}(\Pi_k)} = 0$. Then, the total Augmented Power Dissipation in the system $W_\mathcal{N} + W_\mathcal{L}$ is positive for any non-zero voltage signal whenever for every $k \in \mathcal{L}$ one has $G_{\Pi_k}(\omega) > Y^{\max}_k(\omega) $.
\begin{proof}
 In accordance to lemmas \ref{lemma:Ybound} and \ref{lemma:NetworkPower}, the total Augmented Power Dissipation of the system can be decomposed as
 \begin{align}
  W_\mathcal{N} + W_\mathcal{L} &\geq \sum_{k \in \mathcal{L}}W_{\Pi_k} - y_k |v_k|^2 \nonumber\\
  & \geq \sum_{k \in \mathcal{L}}|v_k|^2\left(G_{\Pi_k} -  Y^{\max}_k\right) > 0.
 \end{align}
\end{proof}
\end{theorem}


\section{Stability Certificates}\label{net}

\subsection{General Network Stability Guidelines}
In order to apply the certificates to actual networks appearing in a microgrid context, we define simple criteria that guarantee existence of paths with sufficiently high Augmented Conductance to support a given load below its crossover frequency $\Omega_c$. In accordance to \eqref{eq:PathConductance}, the path has a high conductance only if all of its elements are conductive enough. On the other hand, as discussed in remarks \ref{remark:Inductive} and \ref{remark:Capacitive}, the Augmented Conductance of inductive elements decreases at high frequencies, and paths with only inductive elements cannot be used to certify stability at high frequencies. At the same time, the Augmented Conductance of capacitive elements goes to zero at small frequencies and grows at high frequencies, and can be naturally used to certify the stability there. Hence, we conclude that if the load is not passive at high frequencies, namely that real part of load admittance remains negative, at least two kinds of paths need to be utilized to certify the overall stability of the system. In the following, we first discuss the properties of inductive and then capacitive paths.

In real microgrids, inductive lines are commonly used to interconnect the loads with sources. The common assumption employed in previous studies on the subject is that the time-constant of inductive elements is bounded from above: $L_k/r_k < \tau_L$. This time constant is usually determined by the kind of wires used for interconnection. Both the inductance $L_k$ and resistance $r_k$ scale linearly with the length of the wire, so the time-constant depends only on the type of wires. Assume that there is an equivalent circuit allowing a proper load path decomposition such that for a given load $k$ there is an inductive path $\Pi_k $ to the source, and the resistance of the elements along the path is given by $r_{e}$ with $e \in \Pi_k$. Then, the conductance of individual element is bounded from below by $G_e > 1/(r_e + r_e \omega^2 \tau_L^2)$. While for the total conductance for a non-augmented ($\phi(\omega) = 0$) power is given by
\begin{equation}
 G_{\Pi_k} = \frac{1}{R_{\Pi_k}} \frac{1}{1+ (\omega \tau_L)^2},
\end{equation}
where $R_{\Pi_k} = \sum_{e\in\Pi_k} r_e$ is the total resistance of the path $\Pi_k$. For a constant $Y^{\max}_k$, independent of $\omega$, this path certifies stability in the band of frequencies $|\omega| < \Omega^L_{k}$ with
\begin{equation}
 \Omega^L_{k} = \frac{1}{\tau_L}\sqrt{\frac{1}{R_{\Pi_k}Y^{\max}_k}-1}.
 \label{eq:Line_Omega}
\end{equation}
If rather than using the upper bound, there is an explicit dependence of the admittance magnitude on $\omega$ the corresponding frequency band is determined by the solutions of inequality $G_{\Pi_k}(\omega) > |y_k(\omega)|$. Note that augmentation with some positive angle $\phi(\omega)$ could improve the bandwidth; however such augmentation will rotate the admittance of a capacitive element into the left half plane so that its real part becomes negative. Therefore, the capacitive elements will inject Augmented Power which compromises the assumption of the Theorem \ref{thm:final_result} where all the network elements, even those not participating in path decomposition, are required to have positive Augmented Conductance under any admissible rotation.  This issue can be dealt with by constructing additional paths to balance the capacitors Augmented Power injection, however this is beyond the scope of the present manuscript where we will only consider the non-augmented inductive paths. 

Next, we proceed to certification of stability for high frequency region, for which paths going through capacitors to the ground are needed. If we assume that the bound on load admittance stays constant for all the frequency range - a CPL assumption, than the only way to certify stability is to place capacitor directly at the load bus. Indeed, given the decay of inductive conductance with frequency according to \eqref{eq:InductiveG}, any path certifying the stability can not go through any of inductive elements. Hence, under this assumption the only path that can certify stability has to go through a capacitive element directly connected to the load bus. This argument illuminates the necessity of the capacitive input filter that stabilizes the high frequency perturbations in the system. For realistic load models, however, the real part of the load admittance does not stay constant and negative over all frequencies and becomes positive above the crossover frequency, after which the load becomes passive. Therefore, for this case one can use the paths going through inductive elements and the need for placing capacitors directly at the load bus is eliminated.  

\begin{figure}
    \centering
    \includegraphics[width=0.5\linewidth]{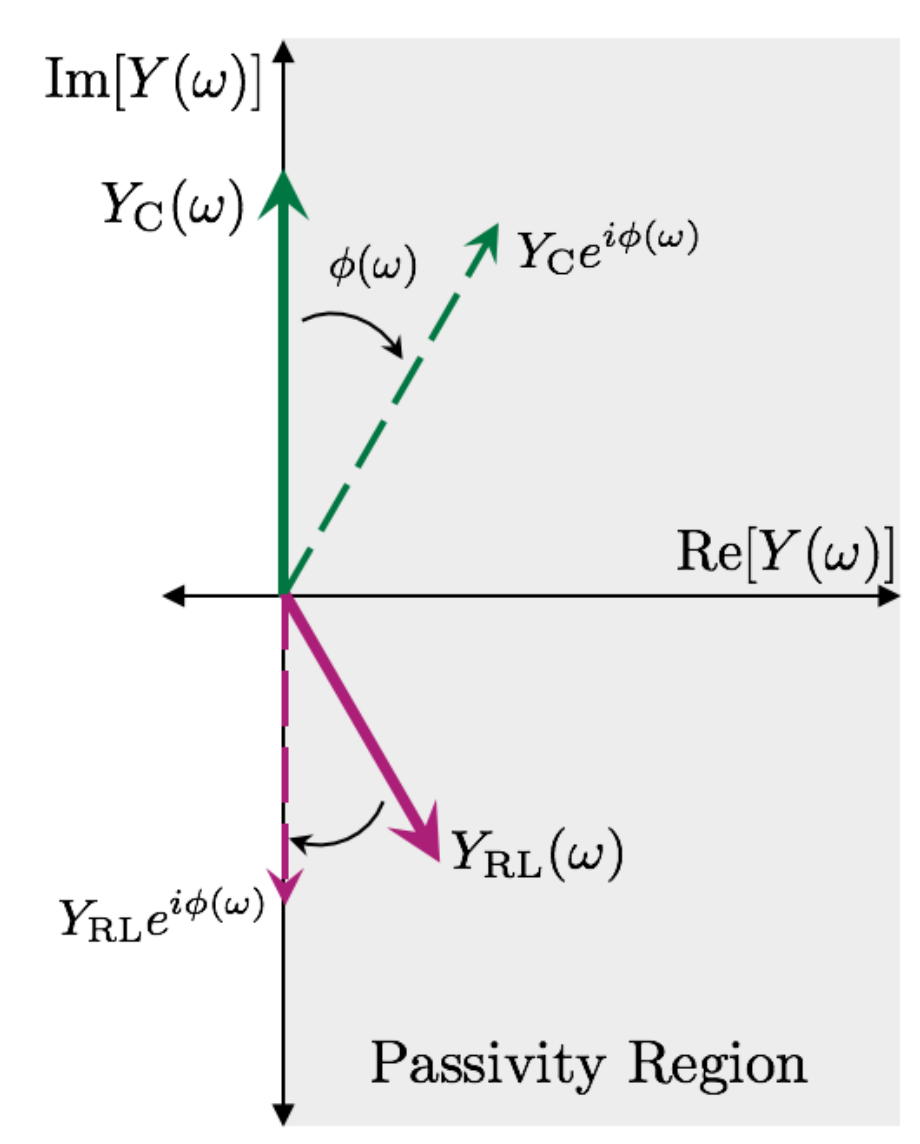}
    \caption{Admittances of a capacitive element, $Y_\text{C}$, and an inductive and resistive element, $Y_{\text{RL}}$ with no augmentation and with an augmentation by a negative rotation of $|\phi_{\text{max}}(\omega)| = \arctan((\omega \tau)^{-1})$. The inductive-resistive elements achieve their maximum allowable dissipation with no rotation while the capacitive elements achieve their maximum allowable dissipation through a negative rotation by $|\phi_{\text{max}}|$}
    \label{fig:Yvectors}
\end{figure}

An Augmented Conduction of a capacitive element is given by $G_{\Pi_k} = -\omega C \sin(\phi(\omega))$. This expression suggests that augmentation with negative $\phi(\omega)$ is essential for a positive Augmented Conductance which can certify stability at high frequencies. As in the previous discussion with inductive lines, there is a maximal angle that is determined by the requirement that all the elements in the network, even those not included in any path, retain positive Augmented Conductance. Given the time-constant $\tau_L$ of inductive elements and using expression \eqref{eq:InductiveG} the maximal negative augmentation angle for a given frequency is given by $\phi(\omega) = -\arctan((\omega \tau_L)^{-1})$. The limitation on the augmentation angle is illustrated by the Figure \ref{fig:Yvectors}. We remind, that this maximum negative augmentation angle can be used only if the capacitive element is directly attached to the load bus. If there is an inductive line between the load and the capacitor the actual augmentation angle could be smaller in absolute value and determined by the effective Augmented Conduction of the whole path. At the maximum negative augmentation angle the conductance of the capacitive path is given by
\begin{equation}
 G_{\Pi_k} = \frac{\omega C_k}{\sqrt{1 + (\omega \tau_L)^2}}
\end{equation}
For constant $Y^{\max}_k$, this capacitor can certify stability for frequencies $|\omega| > \Omega^C_k$ with $\Omega^C_k$ given by
\begin{equation}
 \Omega^C_k = \frac{Y^{\max}_k}{\sqrt{C_k^2- (Y^{\max}_k\tau_L )^2}}
\label{eq:Cap_Omega}
\end{equation}
The overall system can certified to be stable only if for all the loads the two frequency bands given by inductive paths - equation \eqref{eq:Line_Omega}, and capacitive paths - equation \eqref{eq:Cap_Omega} - overlap, i.e. $\Omega_k^C < \Omega_k^L$. The resulting criterion can be formulated either as a condition on $C_k$ or $R_{\Pi_k}$, in the following two equivalent forms:

\begin{equation} \label{eq:cap_constraint}
    C_k > \frac{Y^{\max}_k \tau_L}{\sqrt{1-R_{\Pi_k}Y^{\max}_k}}
\end{equation}
\begin{equation} \label{eq:res_constraint}
    R_{\Pi_k} < \frac{C_k^2-(Y^{\max}_k \tau_L)^2}{Y^{\max}_k C^2}
\end{equation}

\begin{figure}
    \centering
    \includegraphics[width = 0.68\linewidth]{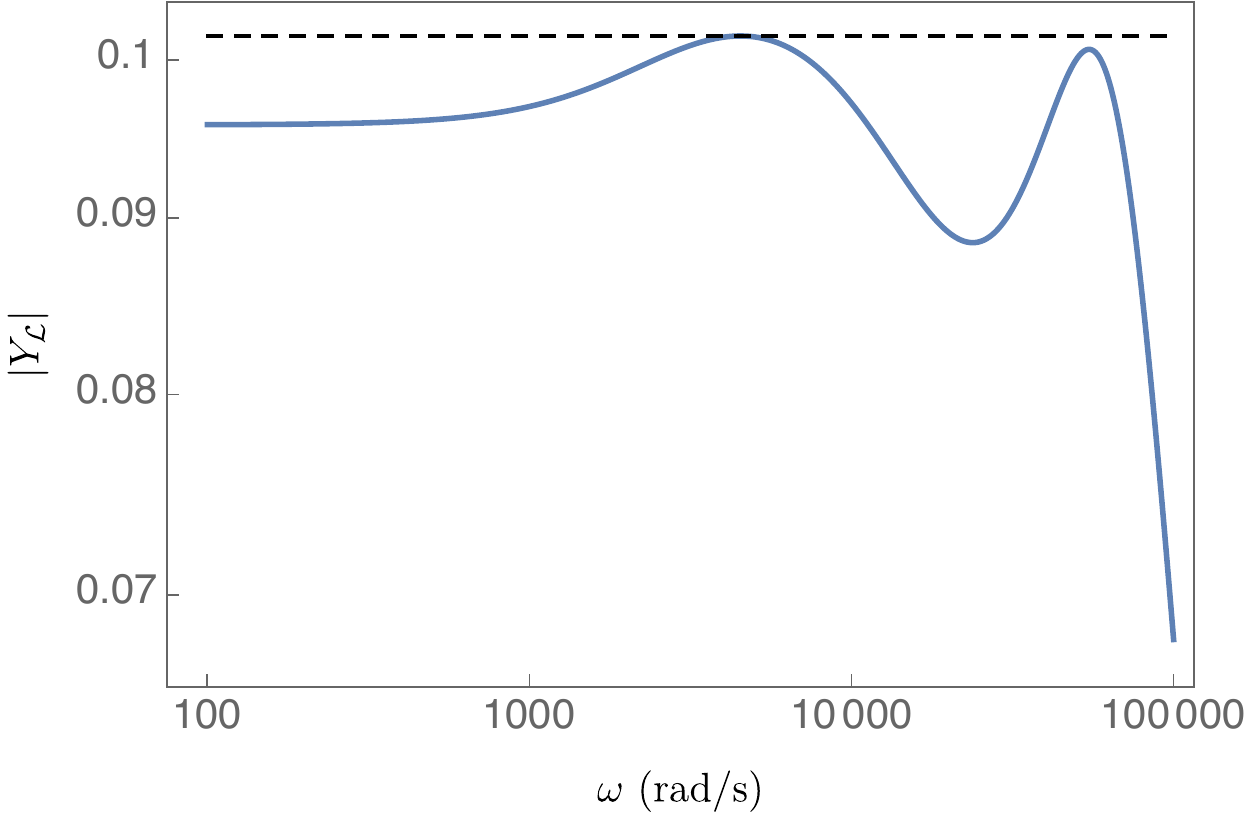}
    \caption{Magnitude of the load admittance transfer function bounded from above by $Y_{\text{max}} \approx 0.1016$}
    \label{fig:BoundingY}
\end{figure}

For an idealized constant power load, the incremental admittance is simply $Y(\omega) = -P/V^2$ where $P$ and $V$ are the equilibrium power consumption and voltage across the load. The load admittance is independent of frequency and therefore the bound on admittance is simply it's magnitude, i.e. $Y^{\max}_k = V^2/P$. Substituting this expression into equation \eqref{eq:cap_constraint} or \eqref{eq:res_constraint} allows for necessary (in the framework of the Augmented Power Dissipation method) stability conditions for the network parameters to be calculated. For more complicated load models or for experimental use, an upper bound on the absolute value of admittance can be evaluated numerically. For example, the magnitude of the admittance transfer function for a simple buck converter is given in Figure \ref{fig:BoundingY} and the upper bound of the admittance magnitude can be discerned graphically. This upper bound is frequency independent; however, unlike with the CPL, the network is only required to be sufficiently conductive for frequencies below the crossover frequency $\Omega_c$. 

\subsection{Relationship to Middlebrook Criterion}

The Middlebrook criterion has been a powerful concept in stability of DC power loads. This criterion is based on the small gain theorem and states that the system is stable if the ratio of the magnitudes of the input impedance of a load to the output impedance of the network is greater than 1, namely 

\begin{equation}
    \frac{|Z_{\text{in, }\mathcal{L}}|}{|Z_{\text{out, }\mathcal{N}}|}>1.
\end{equation}
In practice, the ratio is designed to be much larger than 1. 

For a network path of arbitrary impedance, $Z_{\Pi_k}(\omega) = R(\omega) + j X(\omega)$, the corresponding conductance can be written as

\begin{equation} \label{eq:general_G}
    G_{\Pi_k}(\phi(\omega))=\frac{R(\omega)\cos(\phi(\omega))+X(\omega)\sin(\phi(\omega))}{|Z_{\Pi_k}|^2}.
\end{equation}

Let us now optimize this conductance with and augmentation angle $\phi_{\text{opt}}=\arctan(X/R)$. A rotation by $\phi_{\text{opt}}$ aligns the admittance vector with the real axis. The Augmented Conductance then is simply the magnitude of the admittance given an optimal rotation. We note, that due to the restriction that all elements in the network should remain passive, the optimal rotation can sometimes be not allowable. Figure \ref{fig:Yvectors} shows the admittance vectors of a capacitive element, $Y_C(\omega)$, and an inductive-resistive element, $Y_{RL}(\omega)$. The optimal rotation for the capacitive element is $\phi_{\text{opt}}=-\pi/2$ while the optimal rotation for the inductive-resistive element is $\phi_{\text{opt}}=\arctan(\omega \tau)$. An optimal rotation for either element would cause the other element to exit the passivity region. Therefore, to achieve the maximum allowable Augmented Conductance for the inductive-resistive element, no rotation is performed since a positive rotation which would increase it's maximum Augmented Conductance would make the capacitive element to inject the Augmented Power. However, for the capacitive element, the maximal allowed negative rotation - dictated by an inductive element - is applied as is demonstrated in the figure. After any non-optimal rotation, the resulting vector is in the right-hand side of the complex plane and the resulting Augmenting Conductance is the projection of this rotated vector onto the real axis and can be expressed as

\begin{equation}
G_{\Pi_k}(\phi_{max}(\omega)) = \eta(\phi_{max}(\omega), \omega) |Y|
\end{equation}
where $\eta(\phi(\omega), \omega)$ is an attenuation factor easily derived from equation \eqref{eq:general_G}, varying between 0 and 1, due to the restrictions the passivity requirements impose on the angle $\phi(\omega)$ which may prohibit the optimal rotation. If the optimal rotation is admissible, the attenuation factor is 1. This formulation of the Augmented Conductance allows for Theorem \ref{thm:final_result} to be expressed as 

\begin{align}
    \frac{|Z_{\mathcal{L}}|}{|Z_{\Pi_k}|}>\eta^{-1}(\phi(\omega), \omega)
\end{align}

This echoes the form of the Middlebrook criterion with two noteable differences. First, the impedance ratio must be greater than or equal to 1. In practice, the Middlebrook criterion dictates the load impedance to be much greater than the network impedance. This formulation allows for the quantification of that ratio given network effects. Second, the ratio of concern is that of the load with respect to a specific path, rather than the entire network. The path impedance is bounded from above by the total network impedance. Therefore, even though the impedance ratio is greater than or equal to 1, it is possible to achieve less conservative bounds on the load.   


\section{Numerical Simulation}\label{sec:num}

\subsection{Multi-Microgrid}

\begin{figure}
    \centering
    \includegraphics[width=\linewidth]{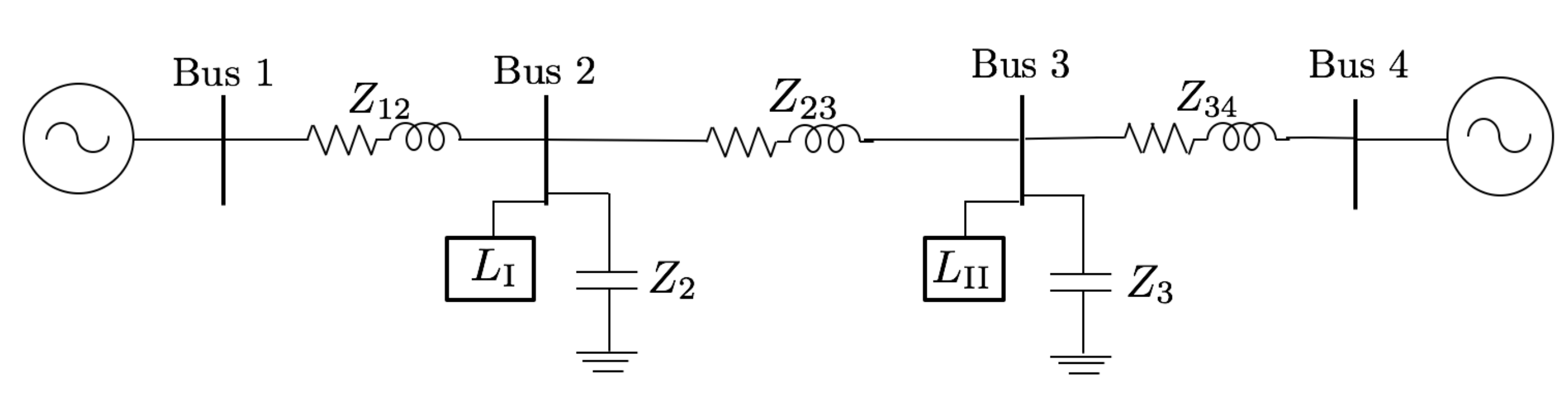}
    \caption{4 bus system representing the interconnection of two microgrids, each comprised of a source and a load.}
    \label{fig:sim_setup}
\end{figure}

To demonstrate the utility of the stability certificates and path decomposition, we consider the interconnection of two microgrids, each represented by a voltage source connected to an aggregated load by an inductive-resitive line as given in Figure \ref{fig:sim_setup}. $L_{\text{I}}$ and $L_{\text{II}}$ are the loads at buses 2 and 3 respectively and are modeled as a constant resistance load interfaced with the network through a buck converter. The converter output voltage is maintained through the addition of a simple lead-lag controller of the form 

\begin{equation}
    G_c(s) = G_{c\infty}\frac{\left(1+\frac{\omega_L}{s}\right) \left(1+\frac{s}{\omega_z}\right)}{\left(1+\frac{s}{\omega_p}\right)}.
\end{equation}

The converter and controller parameters have been drawn from \cite{erickson2007fundamentals} and are defined in Table \ref{tab:sim_param}. The network parameters are given in Table \ref{tab:sim_param}. For simplicity, $L_{\text{I}}$ and $L_{\text{II}}$ are identical with a crossover frequency $\Omega_c \approx 4.8$kHz or $30160$ rad/s and have an admittance magnitude bounded by $Y^{\max}\approx0.1016$, which can be deduced from the model. Therefore, to certify the stability of the system, there must be a path or paths from each load $k$ to a source or ground such that $G_{\Pi_k}(\omega)>Y^{\max}$ for $\omega < \Omega_c$ 

According to discussion in the previous section, stability certification at at low frequencies is done by chosing paths from loads to sources passing through inductive-resistive lines. This can only be done without any augmentation, i.e. $\phi(\omega)=0$ in order to prevent capacitive elements from injecting the Augmented Power. At high frequencies, the paths should go through capacitive elements to the ground and  rotation by a negative non-zero angle is required for capacitors to provide positive Augmented Conductance $G(\omega, \phi(\omega))$. However, the rotation angle is now limited by the condition that the Augmented Conductance of inductive-resisteve elements be at least non-negative which is possible if $|\phi(\omega)|<\arctan((\omega \tau_{max})^{-1})$, where $\tau_{max}$ is the maximum value of $L/r$ for lines in the network. In our case we assume all the lines to have the same $\tau=1 ms$ which plays the role of $\tau_{max}$.

\begin{figure}
    \centering
    \begin{subfigure}[t]{.45\linewidth}
        \centering
        \includegraphics[width=\linewidth]{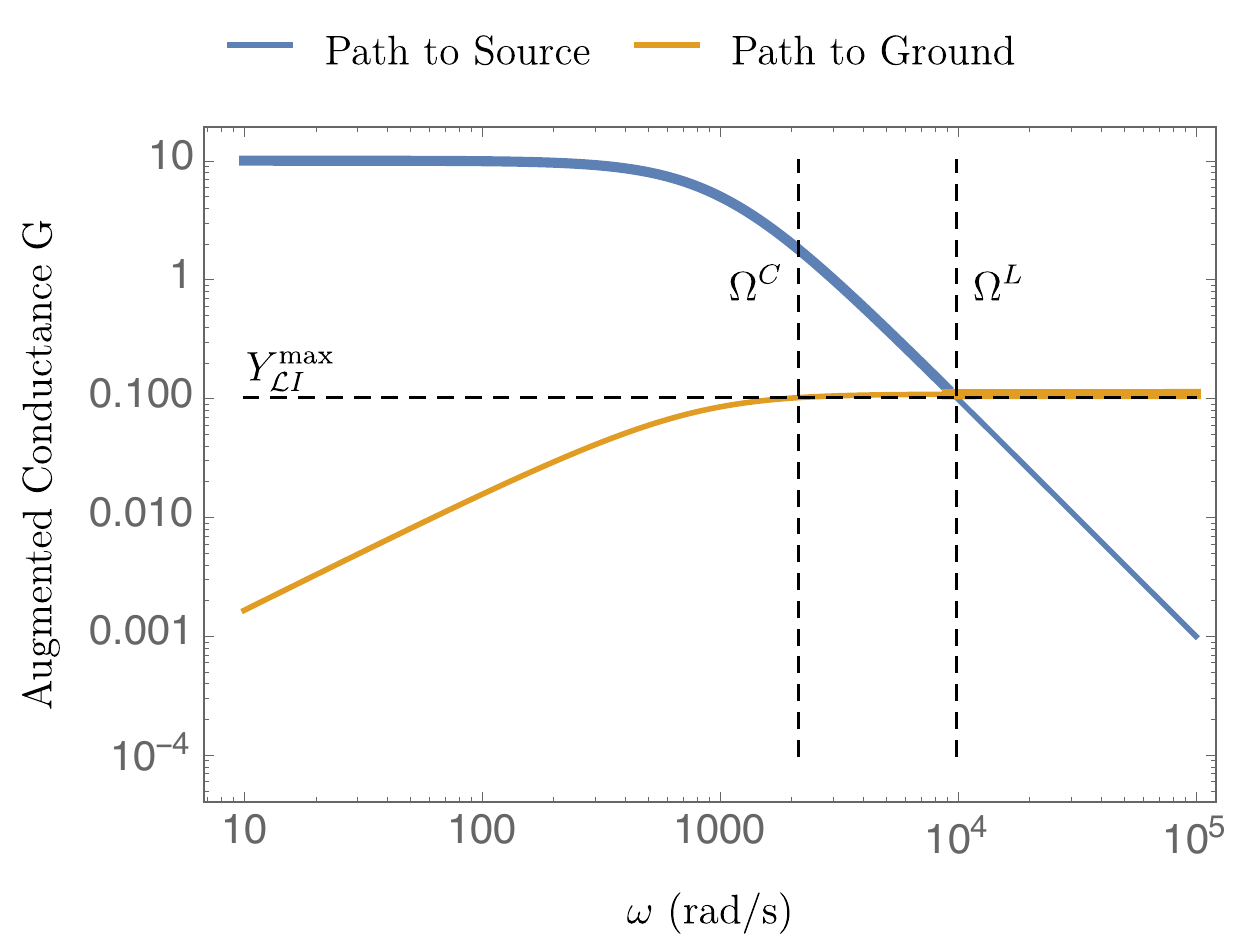}
        \caption{$L_{\text{I}}$}
    \end{subfigure}
    \begin{subfigure}[t]{.45\linewidth}
        \centering
        \includegraphics[width=\linewidth]{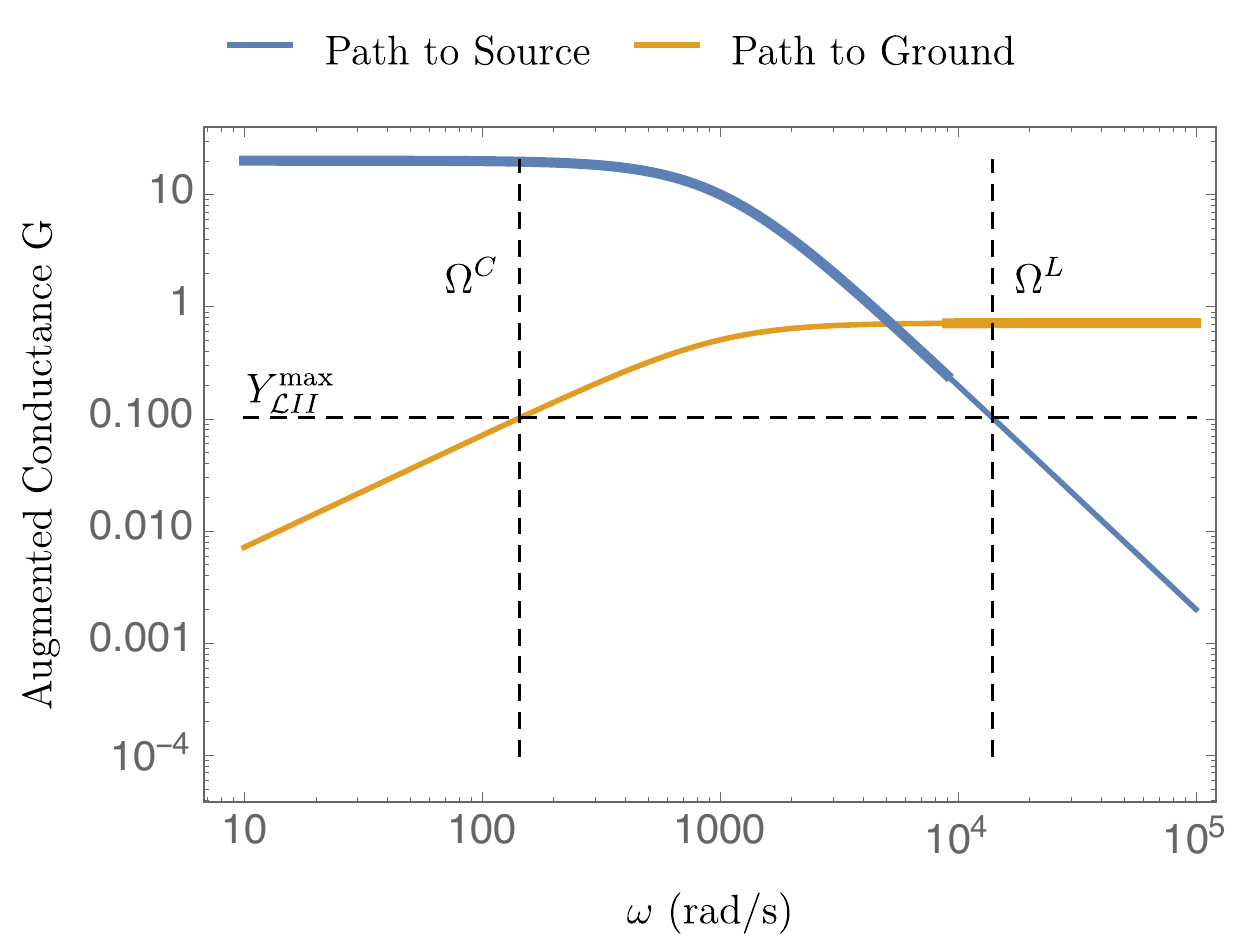}
        \caption{$L_{\text{II}}$}
    \end{subfigure}%
    \caption{Augmented conductance of paths supporting $L_\text{I}$ and $L_\text{II}$ before the addition of $L_{\text{III}}$. The bolded sections represent the overall augmented conductance based on a consistent admissible $\phi(\omega)$.}
    \label{fig:GBefore}
\end{figure}

In the configuration shown by Fig. \ref{fig:sim_setup}, $L_{\text{I}}$ is supported at low frequencies by a path to the source at bus 1 and at high frequencies by the capacitor at bus 2 while $L_{\text{II}}$ is supported at low frequencies by a path to the source at bus 4 and at high frequencies by the capacitor at bus 3. These paths do not overlap and thus no path decomposition is required. The Augmented Conductances of the paths supporting each load are given in Figure \ref{fig:GBefore}. The frequency at which the corresponding path through the line with no augmentation no longer exceeds the load admittance is denoted by $\Omega^L$ and the frequency at which the path through the capacitor, augmented by $\phi = -\arctan((\omega \tau)^{-1})$, begins to exceed the load admittance is denoted by $\Omega^C$. In order to provide the continuous support, these frequency bands should overlap for both of the loads.  The overall Augmented Conductance supporting both loads of example from Fig. \ref{fig:sim_setup} at all the frequency range is shown by the bolded sections of the path Augmented Conductances. A non-zero $\phi$ was chosen to begin at $\omega = 9000$ (rad/s) which is below the crossover frequency of each load but above $\Omega^C$ for each load, the minimum frequency at which the capacitors can certify stability.

\begin{figure}
    \centering
    \begin{subfigure}[b]{\linewidth}
        \centering
        \includegraphics[width=\linewidth]{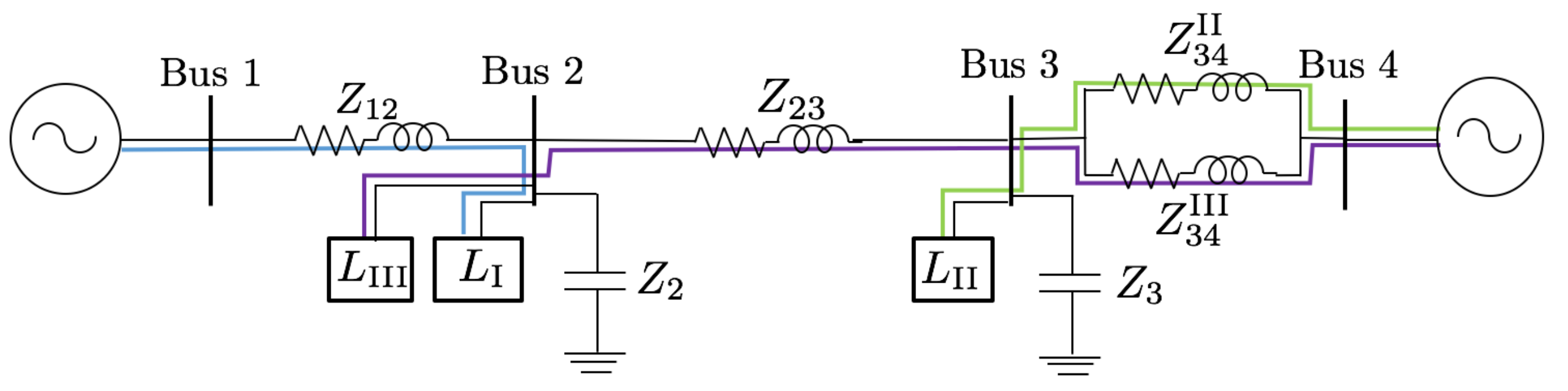}
        \caption{Low Frequency}
        \label{fig:sim_lowf_decomp}
    \end{subfigure}%
    \\
    \begin{subfigure}[b]{\linewidth}
        \centering
        \includegraphics[width=\linewidth]{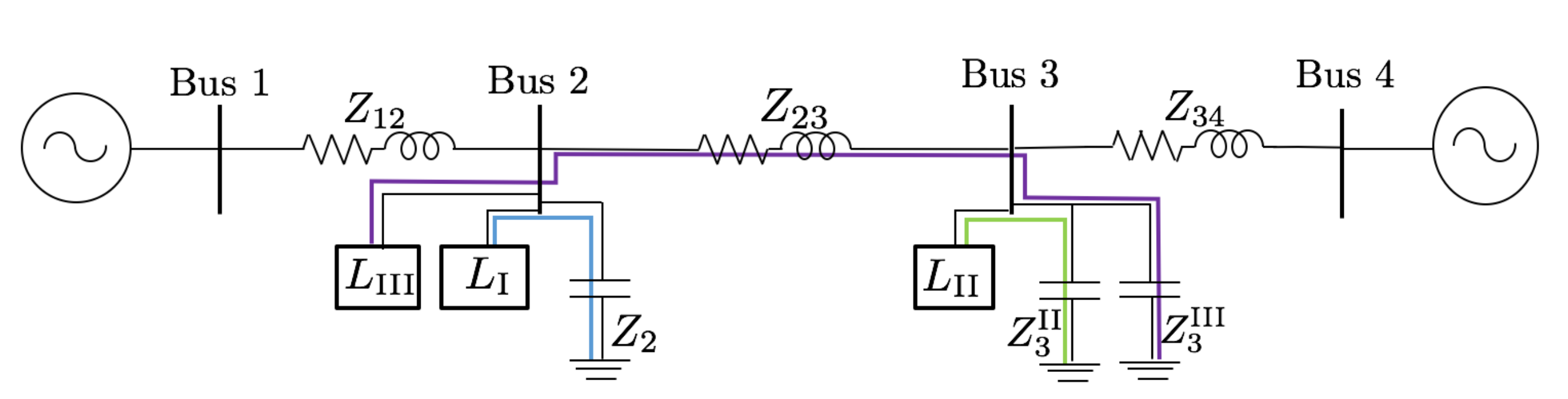}
        \caption{High Frequency}
        \label{fig:sim_highf_decomp}
    \end{subfigure}%
    \caption{High and low frequency path decomposition of the 4 bus multi microgrid system shown in Figure \ref{fig:sim_setup}}
\end{figure}

In the configuration shown by Fig. \ref{fig:sim_setup}, the line connecting buses 2 and 3 is not utilized to support any of the loads. By utilizing this line and it is possible to support additional load. An additional load on bus 2 could be supported through alternate paths such as to the source on bus 4 through the lines connecting buses 2 and 4 for low frequencies and to the capacitor at bus 3 via the line connecting buses 2 and 3 at high frequencies. However, utilizing these paths requires a decomposition of $Z_{34}$ at low frequencies, as in Figure \ref{fig:sim_lowf_decomp}, and $Z_3$ at high frequencies, as in Figure \ref{fig:sim_highf_decomp}. In this decomposition, the superscripts refer to the load being supported by the corresponding network element. 

\begin{figure}
    \centering
    \includegraphics[width=0.75\linewidth]{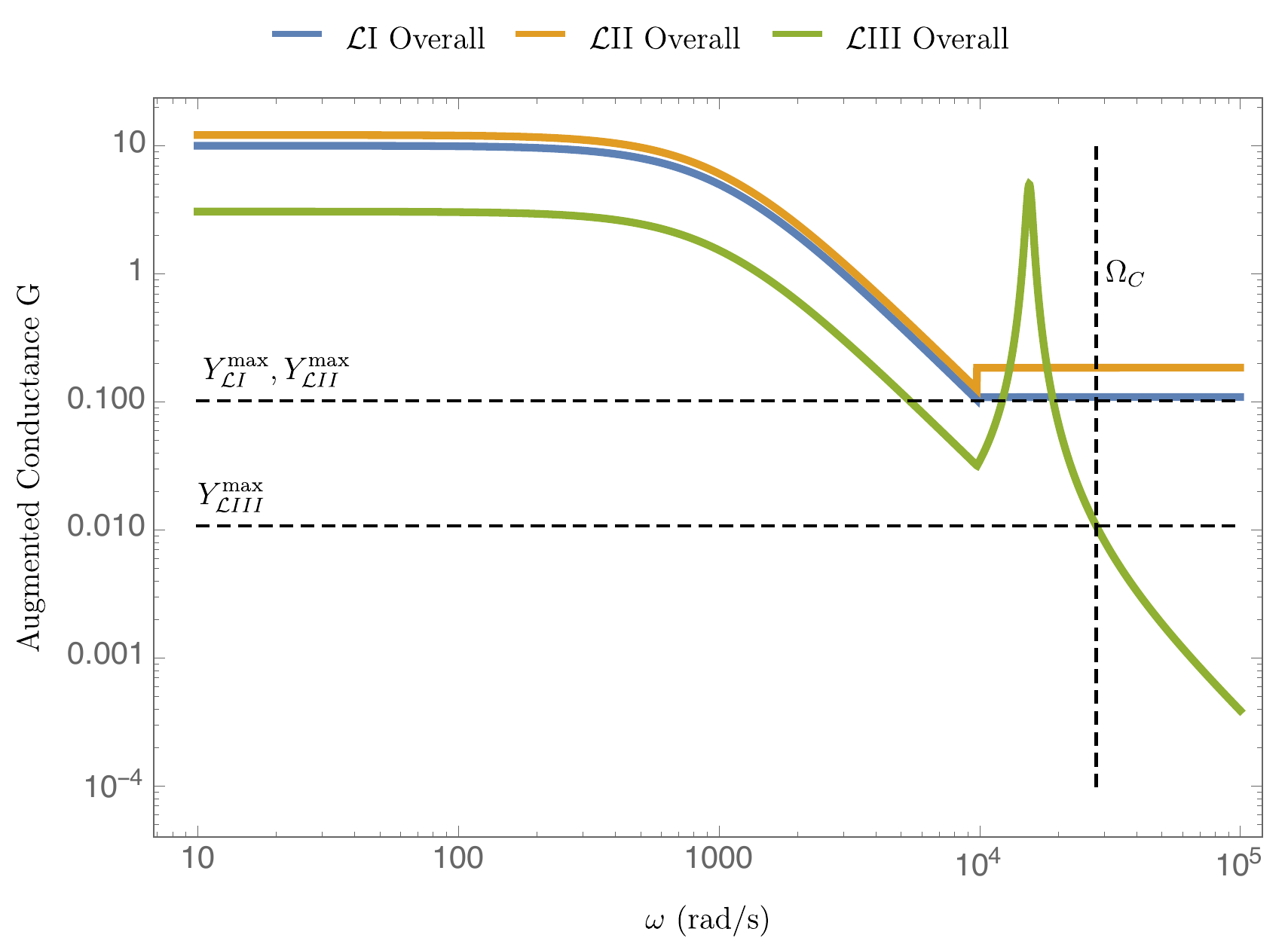}
    \caption{Augmented conductance $G$ of paths supporting $L_\text{I}$, $L_\text{II}$ and  $L_\text{III}$ }
    \label{fig:sim_final_G}
\end{figure}

Figure \ref{fig:sim_final_G} gives the results of an example decomposition of the system and demonstrates how an additional load at bus 2 can be supported while still providing enough Augmented Conduction for original loads. It is important to note that this decomposition is not unique and the maximum additional load, that can be supported is related specifically to the decomposition chosen. Further, the rotation $\phi(\omega)$ that we use to provide support at high frequencies is altered from the maximum allowed because the path containing both the line 23 and the capacitor 3 requires an augmentation different than what gives the maximum for the capacitor, since then the line Augmented Conductance will be zero and an overall path will provide no support. In addition, the rotation chosen must still guarantee that the original loads are still supported. The contribution of the line 23 to the high-frequency path for load 3 can be seen on Fig. \ref{fig:sim_final_G} where the Augmented Conductance for this path starts to decrease with frequency after certain point due to decreasing line conductance. 

The maximum magnitude of load 3 that could be supported is based upon the minimum value of the Augmented Conductance available on it's paths over the frequency range $\omega<\Omega_c$. In Figure \ref{fig:sim_final_G}, the crossover frequency is assumed to be the same for all the loads in the systems and the maximum magnitude that could be supported is $Y_{\mathcal{L}III}^{\text{max}} \approx 0.011$ and the additional decomposition allows for a 10\% increase in loading on bus 2.

\begin{table}[]
    \centering
    \begin{tabular}{|c|c|c|c|}
        \hline
          &Parameter&Description&Value  \\ \hline
          \multirow{5}{*}{\rotatebox[origin = c]{90}{Converter}}&V& Input Voltage&28V\\
          &R& Load Resistance &3 $\Omega$\\
          &C& Capacitance &500 $\mu$F\\
          &L& Inductance &50 $\mu$H\\
          &D&Duty Cycle&0.536\\ \hline
          \multirow{6}{*}{\rotatebox[origin = c]{90}{Controller}}&$G_{c\infty}$& Midband Gain &3.7 \\
          &$w_L$& Lead Zero &500 Hz\\
          &$w_z$& Trailing Zero &1700 Hz\\
          &$w_p$& Pole &14.5 kHz \\
          &H & Sensor Gain & $\frac{1}{3}$\\
          &$V_m$& Voltage of PWM &4\\
          \hline
    \end{tabular}
    \caption{Converter and Control Parameters}
    \label{tab:sim_param}
\end{table}

\begin{table}[]
    \centering
    \begin{tabular}{|c|c|c|}
        \hline
          Parameter&Description&Value  \\ \hline
           $l_{12}$&Line Length between Buses 1 and 2 & 0.5 km\\
           $C_{2}$& Capacitance at Bus 2 & 408.5 $\mu$F\\
           $l_{23}$& Line Length between Buses 2 and 3 & 1.0 km\\
           $C_{3}$& Capacitance at Bus 3&  713.1 $\mu$F\\
           $l_{34}$ & Line Length between Buses 3 and 4& 0.25 km \\
           r& Line Resistance& 0.2 $\Omega$/km\\
           $\tau$ & Time Constant of All Lines & 1 ms\\
          \hline
    \end{tabular}
    \caption{Multi-Microgrid Network Parameters}
    \label{tab:sim_param}
\end{table}

\subsection{Capacitor Placement}
\begin{figure}
    \centering
    \begin{subfigure}[t]{0.5\linewidth}
        \centering
        \includegraphics[width=\linewidth]{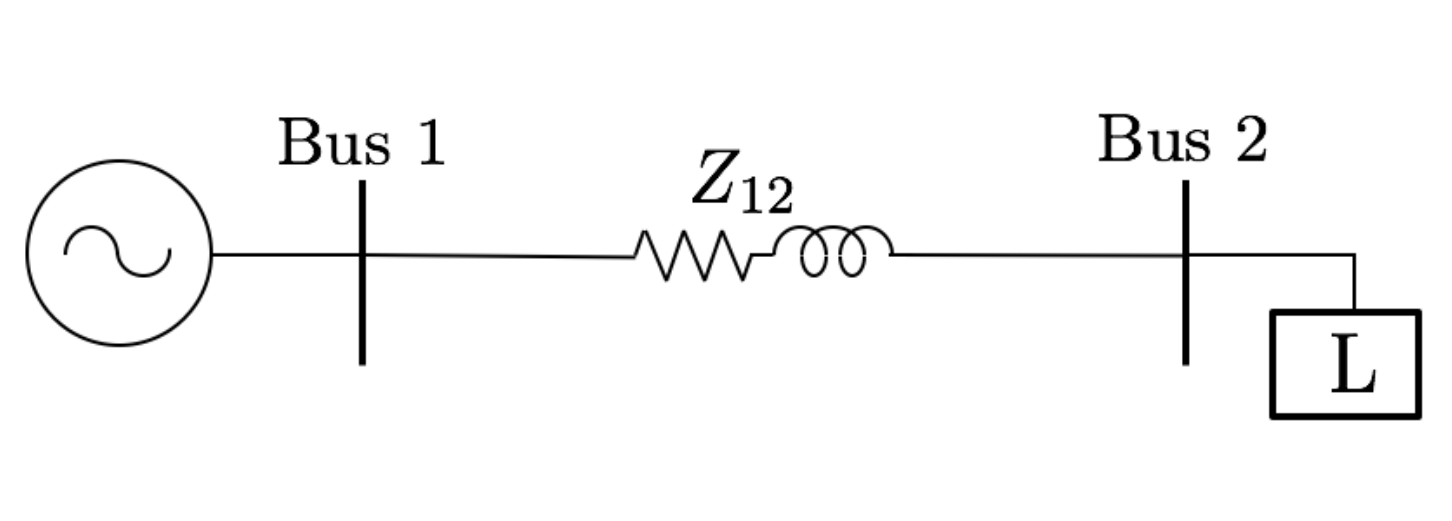}
        \caption{Original System}
    \end{subfigure}%
    \begin{subfigure}[t]{0.5\linewidth}
        \centering
        \includegraphics[width=\linewidth]{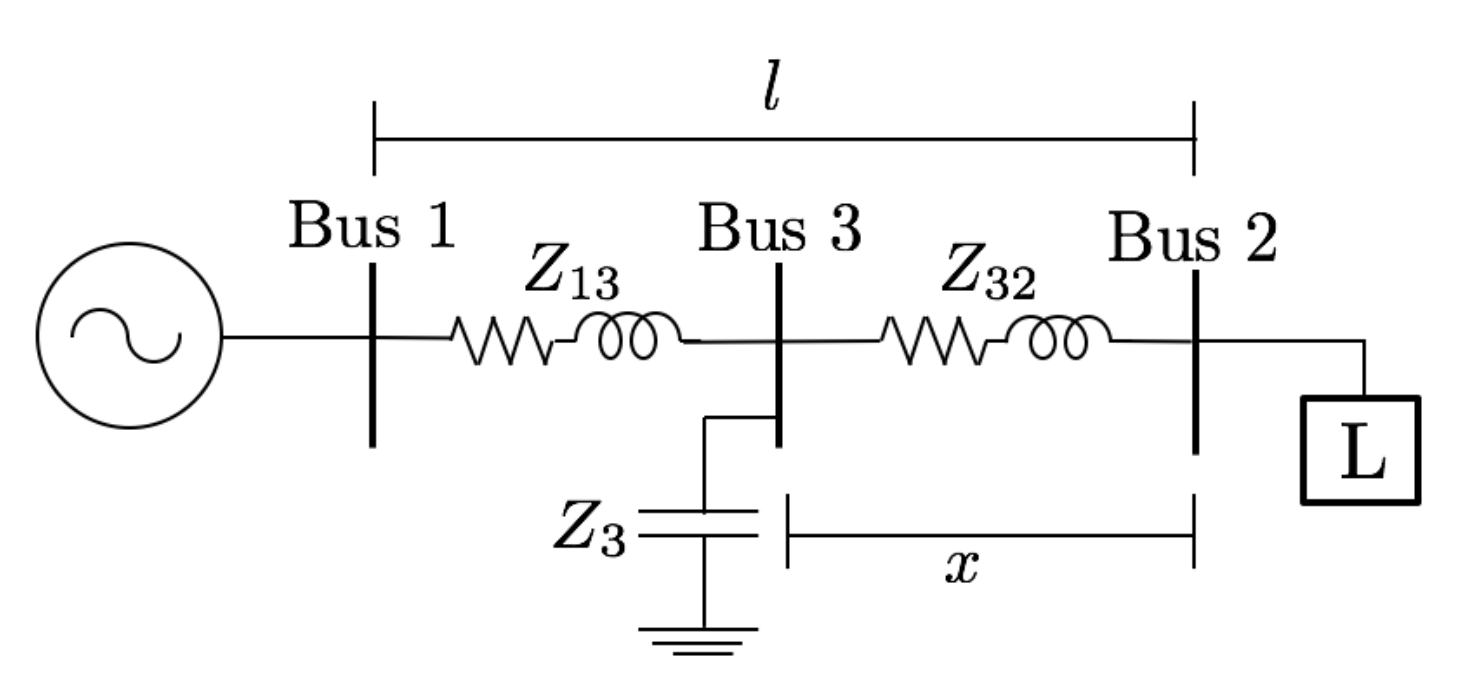}
        \caption{System with addition of stabilizing capacitor}
    \end{subfigure}%
    \caption{2 bus system for capacitor placement}
    \label{fig:sim_cap_placement_sys}
\end{figure}

 These stability certificates do not assume nor require a capacitor to be present at each load bus but rather provide guidance on the placement of capacitors that can support a variety of loads. To illustrate this, we will consider a simple two bus system, given in Figure \ref{fig:sim_cap_placement_sys},  comprised of a constant voltage source and a buck converter load connected through a line, with converter parameters and control settings remain the same as in the previous example. While this system is simple, the methodology utilized in this analysis can easily be replicated for more complex network topologies.

The load becomes passive for $\omega > \Omega_c \approx 4.8$kHz. In the absence of a capacitor in the network, the system stability can be certified by the line connecting buses 1 and 2 for $\omega \leq \Omega^L = 337.17$ Hz, as derived in Equation \eqref{eq:Line_Omega}. The addition of a capacitor must therefore certify the stability of the system for $\Omega^L<\omega<\Omega_c$ via a path to the ground. The simplest case is the placement of the capacitor at bus 2 such that the path to certify stability consists only of the added capacitor. In this case, the there are no inductive elements to diminish the Augmented Conductance at high frequencies and the required capacitance is given by Equation \eqref{eq:cap_constraint} and $C > 96 \mu F$.  

However, the use of Augmented Dissipation and the consideration of the load crossover frequency allows for the certification of the stability of networks where the loads and capacitors are not co-located. In this case, the path to ground consists of the line connecting buses 2 and 3 and the capacitor at bus 3 and has a total impedance of $Z_\Pi = R_{23}+j(\omega L_{23}-(\omega C)^{-1})$. The corresponding augmented conductance is 

\begin{equation}
    G_\Pi(\phi(\omega)) = \frac{R \cos(\phi(\omega)) + \left(\omega L - \frac{1}{\omega C}\right)\sin(\phi(\omega))}{|Z_\Pi|^2}. 
\end{equation}

To adhere to the requirement that all elements remain passive, the rotation $\phi(\omega)$ is limited to the range $-\arctan(\omega \tau)\geq\phi\leq 0$, just as in the previous example since both cases contain only resistive-inductive elements or capacitive elements. Unlike the previous example, since this path contains inductive and capacitive elements the rotation $\phi$ that results in the largest Augmented Conductance changes with $\omega$. At low frequencies, the maximum Augmented Conductance is limited by the line. At high frequencies, the optimal Augmented Conductance would be achieved by a positive $\phi$. However, this rotation would cause the contribution from the capacitive elements to become negative and thus is not allowed. Therefore, at high frequencies, the largest allowable Augmented Conductance is found with $\phi=0$.  

\begin{figure}
    \centering
    \includegraphics[width = 0.75\linewidth]{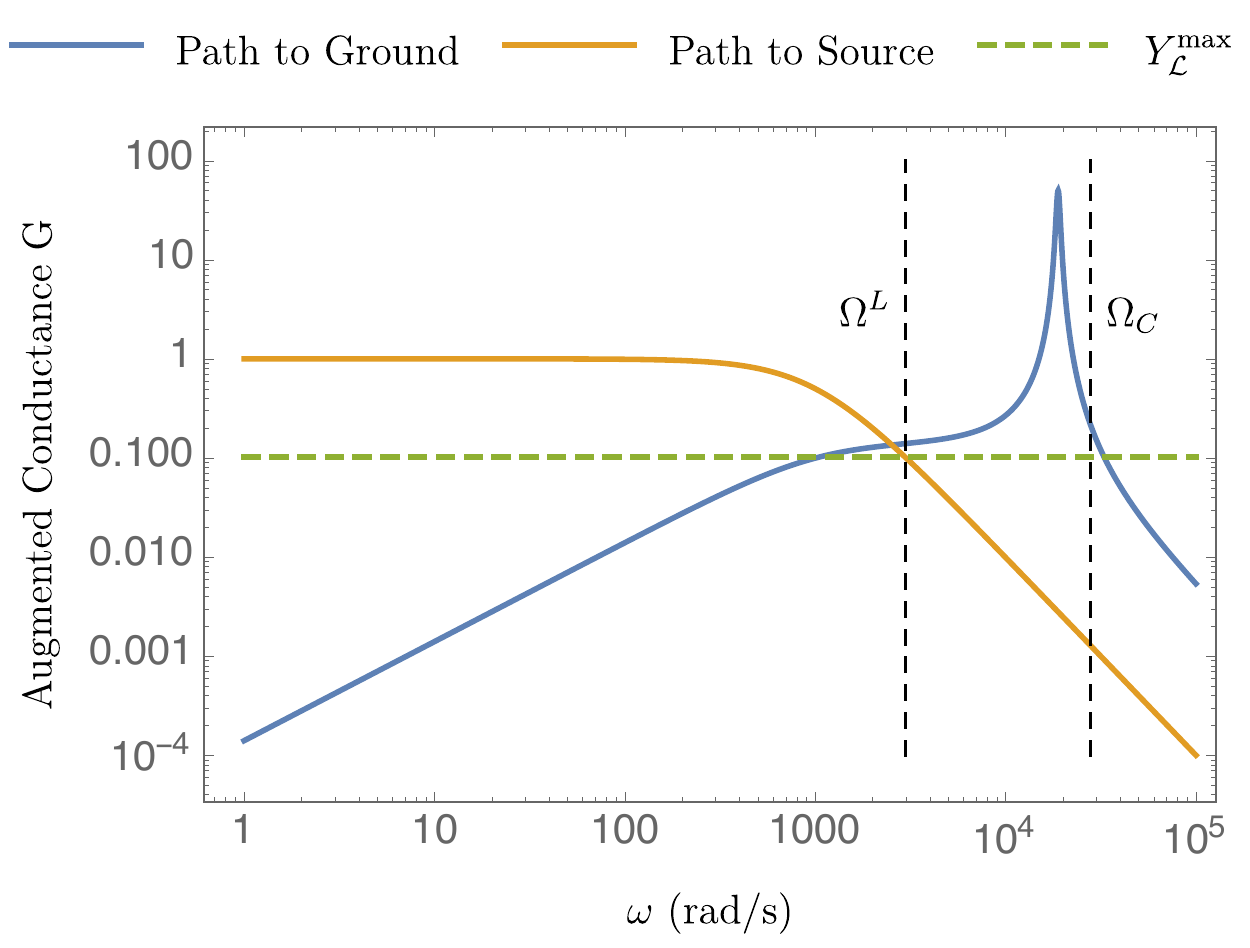}
    \caption{Augmented Conductance $G$ of paths to the source and ground given a capacitor placed at $x/l = 1/50$. }
    \label{fig:G_cap_placement}
\end{figure}

Figure \ref{fig:G_cap_placement} shows the Augmented Conductance of the path through the full line to the source and the path through a segment of the line and the capacitor to the ground with $x = 0.1$km and $C=140\mu$F. For all frequencies below the crossover frequency, $\Omega_c$, one of these paths has an Augmented Conductance that exceeds the maximum load admittance magnitude and therefore stability is guaranteed. It is of note that there are no paths that certify system stability at frequencies slightly above $\Omega_C$, however, since the system is passive in this region there is no requirement for stabilization. As in the previous example, this is an example decomposition and is not intended to present the minimum acceptable capacitance. 

The capacitor could be placed away from the load without compromising stability because a buck converter becomes passive at large frequencies mitigating the concern surrounding decaying conductance of the path to the capacitor due to the inductive elements in the lines. For loads that do not become passive at any frequency, such as ideal constant power loads, this methodology dictates that a capacitor must be placed directly at the load bus. This placement ensures that the Augmented Conductance does not tend toward 0 at very high frequencies due to passing through the inductive element. By considering the magnitude of the transfer function and the crossover frequency of a buck converter rather than approximating it with a constant power load, a greater variety of network topologies can be certified to be stable.

\begin{table}[]
    \centering
    \begin{tabular}{|c|c|c|}
        \hline
          Parameter&Description&Value  \\ \hline
          $l$& Line Length between Buses 1 and 2&5 km\\
          r& Resistance per km & 0.2 $\Omega /$km\\
          $\tau$& Line Time Constant & 1 ms\\
          \hline
    \end{tabular}
    \caption{Capacitor Placement Network Parameters}
    \label{tab:sim_param}
\end{table}

\section{Conclusion}\label{sec:con}

DC microgrids are a powerful and practical option for electrification of remote areas that the main grid cannot reach due to logistical or financial limitations. In this paper, we present a novel method for certifying the stability of a DC network with loads represented by generic transfer functions. While our approach is conceptually similar to the traditional Middlebrook and other similar criteria, it's unique in its ability to capture the effects of network topology in terms of a simple path decomposition. The sufficient stability condition derived in this work relates the Augmented Conductances of the load to the conductance of a path connecting the load bus to some source or ground terminal. For practical purposes, we have identified simple constraints on the network, given in Equations \eqref{eq:cap_constraint} and \eqref{eq:res_constraint}, and presented two simple applications of this theory. 

The proposed approach establishes a new link between the power system stability and classical network flows. This link opens up a path for applying the powerful algorithms developed in computer science and optimization theory to the practical problems of DC microgrid design and analysis. It naturally allows to formulate many practically-relevant problems like stability constrained design and operation in a tractable mathematical form. 

Beside these natural formulations, we believe that the proposed framework can play an important role in establishing universal specifications that would allow for independent decision-making in designing the component and system level controllers, and the network structure. In future, these specifications can form the foundation of true ad-hoc/plug-and-play power systems having the degree of flexibility comparable to modern communication and transportation systems. 

There are many open questions that need to be addressed before these high level goals are reached. One of the most critical, in our opinion is the extension of the proposed framework to nonlinear stability. Power systems lack global stability, and may collapse after strong enough disturbance and linear stability is not sufficient to guarantee reliable and secure operation. Traditional approaches to certification of transient stability of power systems have relied on construction of Lyapunov functions which may not be appropriate for DC microgrid setting. Indeed, as was thoroughly discussed in our work, that load-side converters operating as constant power loads at low frequency naturally destabilize the system. These converters are usually designed by different vendors and their controllers tuned individually and generally not known to the designers and operators of the microgrids. This represents an important limiting factor that largely eliminates the possibility of constructing universal Lyapunov functions that could be used for stability certification. Instead, the more appropriate way to tackle this problem would be to extend the frequency-domain approaches developed in this work to nonlinear settings. This can be naturally accomplished with the help of the Integral Quadratic Constraints framework \cite{megretski1997system} that naturally generalizes some of the key mathematical constructs utilized in our work to the nonlinear setting. 




\bibliographystyle{ieeetr}
\bibliography{mainbib}

\end{document}